\documentclass[preprint,aps,prl,superscriptaddress,amssymb,showpacs]{revtex4}
\usepackage[dvips,final]{graphicx}
\begin{document}
\title{$\pi$-Electron Ferromagnetism in Metal Free Carbon Probed by Soft X-Ray Dichroism}
\author{H. Ohldag}
\email{hohldag@stanford.edu}
\affiliation{Stanford Synchrotron Radiation Laboratory, Stanford University, P.O. Box 20450, Menlo Park, CA 94025, USA}
\author{T. Tyliszczak}
\affiliation{Advanced Light Source, Lawrence Berkeley National Laboratory, Berkeley, CA 94720, USA}
\author{R. H\"{o}hne}
\author{D. Spemann}
\author{P. Esquinazi}
\author{M. Ungureanu}
\author{T. Butz} \affiliation{Institut f\"{u}r Experimentelle
Physik II, Universit\"{a}t Leipzig, Linn\'{e}stra{\ss}e 5, D-04103
Leipzig, Germany} 
\begin{abstract}
Elemental carbon represents a fundamental building block of matter and the possibility of ferromagnetic order in carbon
attracted widespread attention. However, the origin of magnetic order in such a light element is only poorly understood
and has puzzled researchers. We present a spectromicroscopy study at room temperature of proton irradiated metal free
carbon using the elemental and chemical specificity of x-ray magnetic circular dichroism (XMCD). We demonstrate that
the magnetic order in the investigated system originates only from the carbon $\pi$-electron system.
\end{abstract}
 \maketitle
The magnetic properties of carbon have been intensively studied in geology and cosmology \cite{coey:02}, biochemistry
\cite{coronado:00}, physics \cite{odom:00} and material sciences \cite{knisin:04}. The particular interest among the
biophysical and chemical sciences is triggered by the fact that the spin dependent part of the electronic wave function
and its symmetry conditions can have a dramatic influence on the occurrence of certain bonding scenarios
\cite{hoffmann:72}. For example, a magnetic field induced shift in the bonding probability has been used as an argument
for selective diamond formation \cite{little:04}. More general, the possibility of intrinsic long range magnetic order
in carbon is intriguing from a fundamental scientific point of view because its existence in systems containing only s-
and p-electrons demonstrates the importance of correlation effects between these electrons, a scenario that has been
generally neglected so far. However, a conceivable and commonly agreed upon origin of the long range magnetic order in
carbon is still elusive despite the broad interest and extensive theoretical work \cite{Carbonbook}. Reports of
ferromagnetic impurities in nominally impurity-free samples have further increased the scepticism on the existence of
magnetic order in pure carbon systems \cite{hoehne:02,spemann:03}. Altogether it is evident that the study of the
electronic structure of magnetic carbon is required to reveal which electrons contribute to the long range magnetic
order and to elucidate the origin of magnetic order in carbon.


\begin{figure}
\begin{center}
\includegraphics[width=75mm]{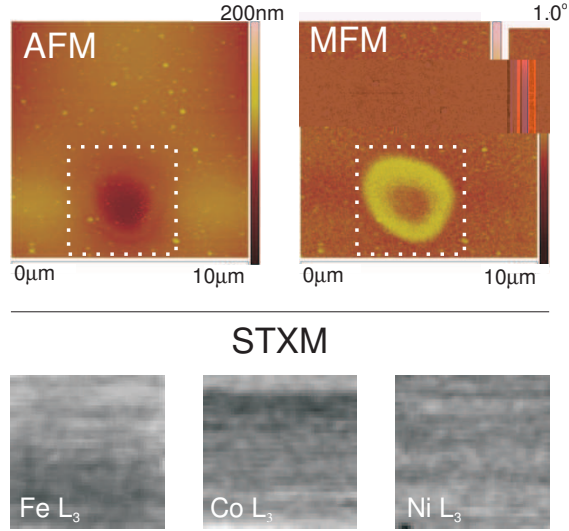}
\caption[]{Top: AFM  and MFM images of a spot irradiated with a 2.25~MeV proton beam and a fluence of 50~nC/$\mu$m$^2$
(sample A). The field of view is 10~$\mu$m. The AFM image reveals the beam impact area. A line scan through its center
reveals a deepening of about 70~nm depth in 5$\mu$m distance. The MFM image suggests a magnetic ``ring" around the
impact area. The measurements were done at ambient conditions using a low moment tip and without applying any magnetic
field. Bottom: STXM images at the Fe, Co and Ni absorption resonance obtained from the area marked with a dotted line
in the force microscopy images above. No contamination is found within the impact area.} \label{fig:sample}
\end{center}
\end{figure}
In this paper we present spectromicroscopic results obtained from micrometer sized magnetic spots that were produced on
thin carbon films by a focused proton beam \cite{esquinazi:02,esquinazi:06}. The approach to ``make carbon magnetic"
through irradiation with a focussed particle beam presents the advantage that the magnetism is confined to only certain
areas and it enables us to directly compare nominally ``magnetic" and ``non-magnetic" carbon in a microscope. For this
purpose we used the scanning transmission x-ray microscope (STXM) located at the elliptical polarizing undulator
beamline 11.0.2 at the Advanced Light Source \cite{kilcoyne:03} in Berkeley, California (USA). This x-ray source
provides intensive soft x-ray beams with variable polarization. The STXM uses a Fresnel Zone plate to focus the
incoming soft x-ray beam to a spot of about 50~nm onto the sample in normal incidence. The x-ray transmission within
the focus area is then detected using a photo diode or photon multiplier behind the sample. By scanning the sample
perpendicular to the optical axis one obtains an image of the lateral distribution of the x-ray absorption cross
section. All measurements of this study were performed at room temperature.

Soft x-ray absorption microscopy makes it possible to obtain element specific information in a complex sample by tuning
the photon energy of the x-rays to the core level absorption resonance of each element. The exact shape and intensity
of such an absorption resonance depends strongly on the local electronic structure of the investigated species and
composition of the sample \cite{stohrbook}. In addition, the polarization dependence of the absorption resonance
(dichroism) carries information about the magnetic order. For example, the transmitted intensity of circular polarized
x-rays depends on the angle between the magnetic moment $\vec{M}$ and the sign of the circular polarization
$\vec{\sigma}$ of the incoming x-rays. This effect is referred to as x-ray magnetic circular dichroism (XMCD) and is
used to quantify the magnetic moment of different elements in a sample. A characteristic feature of XMCD, in contrast
to other non-magnetic forms of circular dichroism, is the fact that the x-ray absorption changes, i.e. the image
contrast is reversed, upon switching the sign of the x-ray polarization or the magnetization and can hence be used in
an x-ray microscope to image the magnetic domain structure of an unknown sample \cite{stohr:93}. We note that while
XMCD at the L-absorption resonances of transition metals can be used to determine the spin and orbital moment of the
element separately this is not possible for K-absorption resonances as in the case of Carbon. The XMCD effect at the
Carbon K-edge only probes the orbital moment \cite{thole:92}, as has been observed in, for example, Fe/C multilayers
\cite{mertins:04}.


For our studies we prepared two thin carbon films  of $\approx$200~nm thickness by pulsed laser deposition (PLD) onto
self-supported, 200~nm thick Si$_3$N$_4$ window for sample A and 100~nm thick for sample B. To obtain graphitic-like
films sample B was deposited at a temperature of 560$^\circ$C with a deposition time of 45~min, while sample A was
prepared at 30$^\circ$C in 30~min to grow a disordered carbon film. The samples were then irradiated by 2.25~MeV
protons using a focussed proton beam to produce an array of magnetic spots with different fluences ranging between 0.1
and 50~nC/$\mu$m$^2$. The spots were separated by 20~$\mu$m each, similar to that produced on oriented graphite
surfaces \cite{han:03}. Proton Induced X-ray Emission (PIXE) measurements performed during irradiation provides upper
limits for the concentration of magnetic ions: for sample A~(B)~$<$~9~(23)~ppm~Fe, and $<$~20~(50)~ppm~Ni. Other
magnetic-contaminant concentrations were below the minimum detection limit of a few ppm for these thin films. The
irradiated areas and surroundings were then characterized by atomic force (AFM) and magnetic force microscopy (MFM) at
room temperature. Fig.~\ref{fig:sample} shows both images for a spot irradiated at 50~nC/$\mu$m$^2$. The AFM image on
the left shows the topography of the sample, while the image on the right shows the phase (magnetic) contrast. One can
identify the beam impact area in the middle of the spot. This area does not show any contrast in the MFM images, while
the ring shaped area surrounding the spot of impact does. Altogether MFM suggests that magnetic order is induced by the
proton irradiation in the area surrounding the beam impact area for the used fluence. Similar contrasts were observed
in irradiated spots in highly oriented pyrolytic graphite (HOPG) produced at high proton currents, suggesting annealing
effects at the center of the spot \cite{esquinazi:06}. STXM images acquired at the Fe, Co and Ni L resonances of this
particular spot shown in the bottom panel of Fig.~\ref{fig:sample} corroborate the findings of the in-situ impurity
characterization during proton irradiation. We find that the level of contamination is well below 100~ppm and that no
ferromagnetic contaminants were introduced during or after proton irradiation that were not detected by PIXE
characterization.


\begin{figure}
\begin{center}
\includegraphics[width=75mm]{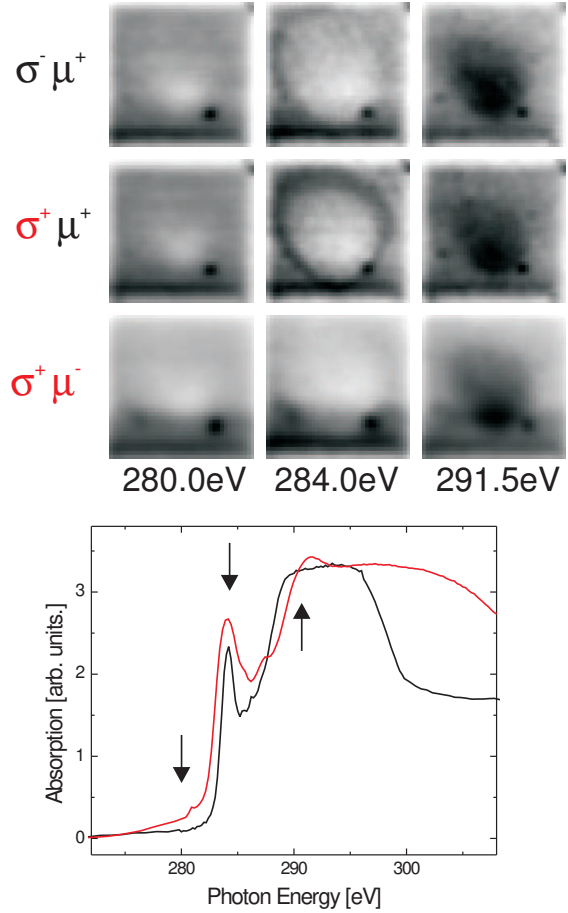}
\caption[]{Carbon K-edge absorption spectrum  (bottom) obtained from the sample prepared at room temperature (black)
and at 560$^\circ$ Celsius substrate temperature (red). The arrows indicate the photon energies for which the STXM
images (top) in the corresponding columns were acquired for a spot irradiated at 50~nC/$\mu$m$^2$ in sample A. The
helicity $(\sigma)$ of the x-rays was reversed between the first and the second row of images. For the third row the
direction $(\mu)$ of the applied field was reversed as well so that both polarization and applied field are opposite to
the situation in the first row. Images acquired at the $\pi^\star$ resonance (284.0~eV) exhibit a clear XMCD signal.}
\label{fig:images}
\end{center}
\end{figure}

We acquired x-ray absorption spectra of both samples as shown on the bottom of Fig.~\ref{fig:images}. The spectra
exhibit a distinct $\pi^\star$ resonance around 284~eV. This resonance is followed by a broad and mostly unstructured
$\sigma^\star$ feature in the case of sample A (black spectra).   The $\sigma^\star$ resonance of sample B however
shows a localized feature around 291.5~eV (red spectra). By comparing our spectra to previous results \cite{stohrbook}
we conclude that the structure of sample A is mostly disordered while the annealing procedure of sample B has led to
partial graphitization causing the more distinct features in XAS. Figure~\ref{fig:images} also shows STXM images
obtained at several photon energies with different orientation of the applied field ($\vec{H}=\pm~600\textrm{Oe}$) and
circular polarization ($\vec{\sigma}=\pm1$). The field was applied parallel to the incident x-rays and perpendicular to
the surface using a permanent magnet. The first column of images was acquired below the resonance at 280~eV, while the
images in the right column were acquired within the $\sigma^\star$ resonance for sample A. We do not observe a
dependence of the image contrast on the polarization of the x-rays or the applied field and therefore find that these
images only provide information about local variation of the film thickness ($h\nu=284$~eV) and chemical bonding
($h\nu=291.5$~eV). The proton irradiation leads to the removal of material and increase in empty $\sigma^\star$ states
in the center of the beam impact area. We can conclude that the $\sigma$ electron system does not contribute to the
magnetic moment since it does not show any XMCD contrast.

The situation is different for images acquired at the $\pi^\star$ resonance at 284.0~eV, see Fig.~\ref{fig:images}.
First the area around the center appears in a light grey but it becomes very distinct and dark upon reversing the
polarization of the x-rays. Subsequent reversal of the direction of the applied field restores the original alignment
of applied field and x-ray polarization and consequently the image exhibits the original contrast again. We conclude
that the observed x-ray magnetic circular dichroism is caused by long range magnetic order originating from the carbon
$\pi$-electron system. We note, that the XMCD effect found here is very small, of the order of $\approx0.1$\%, and
hence superimposed by other non magnetic effects. For this reason we do not observe a distinct black/white contrast as
it is usually the case for bulk transition metal samples \cite{stohr:93} but only a more subtle contrast.

We use the size of the observed dichroism to give  an estimate of the size of the magnetic moment per carbon atom. The
magnetic contrast that we observe is about an order of magnitude smaller than what has been observed at the Fe K-edge
by Sch\"utz et al. applying a similar background subtraction \cite{schutz:87}. Taking into account the different
non-magnetic absorption cross section at the Fe and C K-absorption edge we estimate that the average orbital moment per
Carbon atom in the magnetic ring area is between $5\cdot10^{-4}$ and $1\cdot10^{-3}$ Bohr magneton \cite{thole:92}.
This implicates that the observed XMCD contrast cannot be due to paramagnetic or diamagnetic effects. Were
paramagnetism present at the spot region, it would provide two orders of magnitude smaller contrast at the applied
field of 600~Oe and at room temperature. We note further that the whole sample A is paramagnetic with magnetization at
saturation of $\approx$1~emu/g measured at 2~K in similar carbon films \cite{hoehne:04}. Because of the localized
heating produced by the proton beam we expect actually a decrease of the paramagnetism at the spot center. However,
neither the center of the spot nor the paramagnetic matrix shows  XMCD contrast. We therefore conclude that neither
para- nor diamagnetism is responsible for the XMCD contrast results, taking also into account the SQUID measurements on
irradiated HOPG \cite{esquinazi:02,esquinazi:06} and the similarities observed in MFM between the spots produced in
HOPG \cite{esquinazi:06} and in sample A.

\begin{figure}
\begin{center}
\includegraphics[width=75mm]{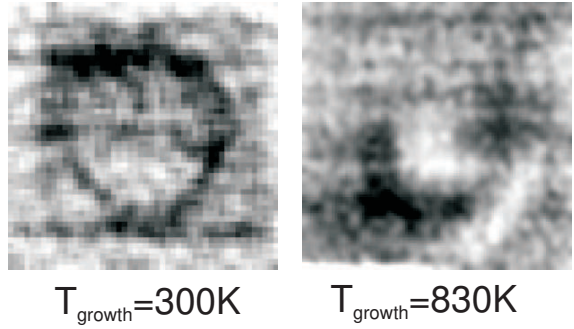}
\caption[]{XMCD image of a spot on sample A irradiated at a fluence of 20~nC/$\mu$m$^2$ is shown on the left. The
center is surrounded by a distinct magnetic ring. The magnetic image on the right was obtained from a spot on sample B
(10~nC/$\mu$m$^2$). Alternating black and white areas indicate that the magnetic area decomposes in different magnetic
domains.} \label{fig:domains}
\end{center}
\end{figure}

We will now address the influence of the partial graphitization on the magnetism of the sample. The previous results
were obtained from sample A that was prepared at room temperature and hence does not show any significant
graphitization. By applying an external magnetic field of 600~Oe it was possible to align the magnetization
perpendicular to the surface (at the ring area of the irradiated spot) so that an homogeneous XMCD effect could be
observed. Sample B is different since it was prepared at 560~$^\circ$C and exhibits a more graphitic character.
Fig.~\ref{fig:domains} shows XMCD images  acquired from spots on sample A and B irradiated with a fluence of 20 and
10~nC/$\mu$m$^2$ respectively. To enhance the XMCD effect and to normalize out other contributions to the image
contrast we first calculated the asymmetry image $I= [I(\pi^\star)-I(\sigma^\star)]/[I(\pi^\star)+I(\sigma^\star)]$ and
then subtracted images acquired with opposite polarization. These images are now freed from other non-magnetic contrast
sources and the magnetic ring can clearly be identified in sample A \cite{footnote1}. The graphitized sample however
does not exhibit a magnetic ring at this fluence  and it appears as if the magnetic moment of the entire spot cannot be
aligned with the external field of 600~Oe. Instead, we find areas exhibiting black and white XMCD contrast around the
beam impact area in the center caused by a residual component of the direction of the magnetic moments perpendicular to
the surface, while most of the magnetic moment prefers to align parallel to the surface and does not cause an XMCD
contrast in the microscope. We conclude that the irradiated spot on the graphitized sample exhibits a preferred
alignment of the magnetization that is parallel to the surface (magnetic anisotropy) compared to the disordered one.
The residual out of plane component of the preferred axis of the magnetization varies around the beam impact area, i.e.
an evidence for the formation of magnetic domains. The fact that we are able to observe a small variation of the
preferred axis of the magnetization without fully magnetizing the spot indicates that the XMCD - and hence its magnetic
moment - observed on the graphitized sample is larger than for the non-graphitized sample.


Using element specific XMCD we have demonstrated that proton irradiation leads to ferromagnetic order in carbon that
originates from the spin-polarization of the carbon $\pi$-electrons.  Both disordered and partly-graphitized films
exhibit an XMCD signal. The main difference between the two samples is that the XMCD signal and hence the magnetic
moment in the partly graphitized sample is larger. In addition the magnetization of the partly-graphitized film
exhibits a distinct preferred magnetic axis parallel to the surface. This observation furthermore corroborates the role
of the $\pi$ electrons in the magnetism of carbon since larger order in the $\pi$ bonds leads to higher value and
anisotropy of the orbital moment and hence anisotropy of the magnetization. Regarding the role played by hydrogen in
the origin of the magnetic order in carbon we note the following. Taking into account that at the used proton energy
only a negligible amount of them are captured in the 200~nm thick films, were hydrogen one of the necessary ingredients
to trigger magnetic order in carbon then it should be the hydrogen already present in the sample, as measured in HOPG
samples \cite{reichart:06}. Hydrogen dissociation plus its bonding at atomic lattice defects, both produced by
irradiation, may then play a role in the observed magnetic order.

We thank Dr.  H. Schmidt for allowing us the use of her MFM apparatus and J. St\"ohr and H.C. Siegmann for comments and
discussions. The work at the SSRL and the ALS is supported by the US Department of Energy, Office of Basic Energy
Sciences. The work in Leipzig is supported by the DFG under DFG ES 86/11-1 and the European Union project
``Ferrocarbon". 
\bibliographystyle{apsrev}

\begin{thebibliography}{21}
\expandafter\ifx\csname natexlab\endcsname\relax\def\natexlab#1{#1}\fi \expandafter\ifx\csname
bibnamefont\endcsname\relax
  \def\bibnamefont#1{#1}\fi
\expandafter\ifx\csname bibfnamefont\endcsname\relax
  \def\bibfnamefont#1{#1}\fi
\expandafter\ifx\csname citenamefont\endcsname\relax
  \def\citenamefont#1{#1}\fi
\expandafter\ifx\csname url\endcsname\relax
  \def\url#1{\texttt{#1}}\fi
\expandafter\ifx\csname urlprefix\endcsname\relax\def\urlprefix{URL }\fi \providecommand{\bibinfo}[2]{#2}
\providecommand{\eprint}[2][]{\url{#2}}

\bibitem[{\citenamefont{et~al.}(2002)}]{coey:02}
\bibinfo{author}{\bibfnamefont{J.~M.~D.~Coey} \bibnamefont{et~al.}},
  \bibinfo{journal}{Nature} \textbf{\bibinfo{volume}{420}},
  \bibinfo{pages}{156} (\bibinfo{year}{2002}).

\bibitem[{\citenamefont{et~al.}(2000{\natexlab{a}})}]{coronado:00}
\bibinfo{author}{\bibfnamefont{E.~Coronado} \bibnamefont{et~al.}},
  \bibinfo{journal}{Nature} \textbf{\bibinfo{volume}{408}},
  \bibinfo{pages}{447} (\bibinfo{year}{2000}{\natexlab{a}}).

\bibitem[{\citenamefont{et~al.}(2000{\natexlab{b}})}]{odom:00}
\bibinfo{author}{\bibfnamefont{T.~W.~Odom} \bibnamefont{et~al.}},
  \bibinfo{journal}{Science} \textbf{\bibinfo{volume}{290}},
  \bibinfo{pages}{1549} (\bibinfo{year}{2000}{\natexlab{b}}).

\bibitem[{\citenamefont{et~al.}(2004{\natexlab{a}})}]{knisin:04}
\bibinfo{author}{\bibfnamefont{L.~Knisin-E.} \bibnamefont{et~al.}},
  \bibinfo{journal}{Nature} \textbf{\bibinfo{volume}{43}}, \bibinfo{pages}{672}
  (\bibinfo{year}{2004}{\natexlab{a}}).

\bibitem[{\citenamefont{Hoffmann and Woodward}(1972)}]{hoffmann:72}
\bibinfo{author}{\bibfnamefont{R.}~\bibnamefont{Hoffmann}} \bibnamefont{and}
  \bibinfo{author}{\bibfnamefont{R.~B.} \bibnamefont{Woodward}},
  \bibinfo{journal}{Chemie in Unserer Zeit} \textbf{\bibinfo{volume}{6}},
  \bibinfo{pages}{167} (\bibinfo{year}{1972}).

\bibitem[{\citenamefont{Little and Goddard}(2004)}]{little:04}
\bibinfo{author}{\bibfnamefont{R.~B.} \bibnamefont{Little}} \bibnamefont{and}
  \bibinfo{author}{\bibfnamefont{R.}~\bibnamefont{Goddard}},
  \bibinfo{journal}{J. Appl. Phys.} \textbf{\bibinfo{volume}{95}},
  \bibinfo{pages}{2702} (\bibinfo{year}{2004}).

\bibitem[{\citenamefont{Makarova and Palacio}(2006)}]{Carbonbook}
\bibinfo{editor}{\bibfnamefont{T.}~\bibnamefont{Makarova}} \bibnamefont{and}
  \bibinfo{editor}{\bibfnamefont{F.}~\bibnamefont{Palacio}}, eds.,
  \emph{\bibinfo{title}{Carbon Based Magnetism}} (\bibinfo{publisher}{Elsevier
  B.V}, \bibinfo{address}{Amsterdam}, \bibinfo{year}{2006}).

\bibitem[{\citenamefont{{H\"{o}hne} and Esquinazi}(2002)}]{hoehne:02}
\bibinfo{author}{\bibfnamefont{R.}~\bibnamefont{{H\"{o}hne}}} \bibnamefont{and}
  \bibinfo{author}{\bibfnamefont{P.}~\bibnamefont{Esquinazi}},
  \bibinfo{journal}{Adv. Mat.} \textbf{\bibinfo{volume}{14}},
  \bibinfo{pages}{753} (\bibinfo{year}{2002}).

\bibitem[{\citenamefont{et~al.}(2003{\natexlab{a}})}]{spemann:03}
\bibinfo{author}{\bibfnamefont{D.~Spemann} \bibnamefont{et~al.}},
  \bibinfo{journal}{Nucl. Inst. Meth. Phys. Res. B.}
  \textbf{\bibinfo{volume}{210}}, \bibinfo{pages}{531}
  (\bibinfo{year}{2003}{\natexlab{a}}).

\bibitem[{\citenamefont{et~al.}(2003{\natexlab{b}})}]{esquinazi:02}
\bibinfo{author}{\bibfnamefont{P.~Esquinazi} \bibnamefont{et~al.}},
  \bibinfo{journal}{Phys. Rev. Lett.} \textbf{\bibinfo{volume}{91}},
  \bibinfo{pages}{227201} (\bibinfo{year}{2003}{\natexlab{b}}).

\bibitem[{\citenamefont{et~al.}(2006{\natexlab{b}})}]{esquinazi:06}
\bibinfo{author}{\bibfnamefont{P.~Esquinazi} \bibnamefont{et~al.}},
in Ref.~7, pp. \bibinfo{pages}{437--462}.

\bibitem[{\citenamefont{Kilcoyne et~al.}(2003)\citenamefont{Kilcoyne,
  Tyliszczak, Steele, Fakra, Hitchcock, Franck, Anderson, and
  Harteneck}}]{kilcoyne:03}
\bibinfo{author}{\bibfnamefont{A.~L.~D.} \bibnamefont{Kilcoyne}} \bibnamefont{et~al.},
  \bibinfo{journal}{J. Synchrotron Radiat.} \textbf{\bibinfo{volume}{10}},
  \bibinfo{pages}{125} (\bibinfo{year}{2003}).

\bibitem[{\citenamefont{{St\"{o}hr}}(1992)}]{stohrbook}
\bibinfo{author}{\bibfnamefont{J.}~\bibnamefont{{St\"{o}hr}}},
  \emph{\bibinfo{title}{NEXAFS Spectroscopy}}, vol.~\bibinfo{volume}{25} of
  \emph{\bibinfo{series}{Springer Series in Surface Sciences}}
  (\bibinfo{publisher}{Springer}, \bibinfo{address}{Heidelberg},
  \bibinfo{year}{1992}).

\bibitem[{\citenamefont{{St\"{o}hr} et~al.}(1993)\citenamefont{{St\"{o}hr}, Wu,
  Hermsmeier, Samant, Harp, Koranda, Dunham, and Tonner}}]{stohr:93}
\bibinfo{author}{\bibfnamefont{J.}~\bibnamefont{{St\"{o}hr}}} \bibnamefont{et~al.},
  \bibinfo{journal}{Science} \textbf{\bibinfo{volume}{259}},
  \bibinfo{pages}{658} (\bibinfo{year}{1993}).

\bibitem[{\citenamefont{Thole et~al.}(1992)\citenamefont{Thole, Carra, Sette,
  and van~der Laan}}]{thole:92}
\bibinfo{author}{\bibfnamefont{B.~T.} \bibnamefont{Thole}},
  \bibinfo{author}{\bibfnamefont{P.}~\bibnamefont{Carra}},
  \bibinfo{author}{\bibfnamefont{F.}~\bibnamefont{Sette}}, \bibnamefont{and}
  \bibinfo{author}{\bibfnamefont{G.}~\bibnamefont{van~der Laan}},
  \bibinfo{journal}{Phys. Rev. Lett.} \textbf{\bibinfo{volume}{68}},
  \bibinfo{pages}{1943} (\bibinfo{year}{1992}).

\bibitem[{\citenamefont{et~al.}(2004{\natexlab{b}})}]{mertins:04}
\bibinfo{author}{\bibfnamefont{H.-C.~Mertins} \bibnamefont{et~al.}},
  \bibinfo{journal}{Europhys. Lett.} \textbf{\bibinfo{volume}{66}},
  \bibinfo{pages}{743} (\bibinfo{year}{2004}{\natexlab{b}}).

\bibitem[{\citenamefont{et~al.}(2003{\natexlab{c}})}]{han:03}
\bibinfo{author}{\bibfnamefont{K.-H.~Han} \bibnamefont{et~al.}},
  \bibinfo{journal}{Adv. Mat.} \textbf{\bibinfo{volume}{15}},
  \bibinfo{pages}{1719} (\bibinfo{year}{2003}{\natexlab{c}}).

\bibitem[{\citenamefont{{Sch\"{u}tz} et~al.}(1987)\citenamefont{{Sch\"{u}tz},
  Wagner, Wilhelm, Kienle, Zeller, Frahm, and Materlik}}]{schutz:87}
\bibinfo{author}{\bibfnamefont{G.}~\bibnamefont{{Sch\"{u}tz}}}  \bibnamefont{et~al.},
  \bibinfo{journal}{Phys. Rev. Lett.} \textbf{\bibinfo{volume}{58}},
  \bibinfo{pages}{737} (\bibinfo{year}{1987}).

\bibitem[{\citenamefont{et~al.}(2004{\natexlab{c}})}]{hoehne:04}
\bibinfo{author}{\bibfnamefont{R.~{H{\"o}hne}.} \bibnamefont{et~al.}},
  \bibinfo{journal}{J. Magn. Magn. Mater.}
  \textbf{\bibinfo{volume}{e839-e840}}, \bibinfo{pages}{272}
  (\bibinfo{year}{2004}{\natexlab{c}}).

\bibitem[{\citenamefont{et~al.}(2004{\natexlab{c}})}]{footnote1}
\bibinfo{misc}{\bibnamefont{The fact that images obtained from different spots with similar spatial dimensions and fluence,
as it is the case here, yield qualitatively identical results clearly shows the reproducibility of our
results.}}

\bibitem[{\citenamefont{Reichart}(2006)}]{reichart:06}
\bibinfo{author}{\bibfnamefont{P.}~\bibnamefont{Reichart}} \bibnamefont{et~al.},
  \bibinfo{journal}{Nucl. Instr and Meth.in Phys. Res. B}
  \textbf{\bibinfo{volume}{249}}, \bibinfo{pages}{286-291}
  (\bibinfo{year}{2006}).

\end{thebibliography}

\end{document}